\documentclass{raa}
\usepackage{graphicx,times}             
\usepackage{natbib}
\usepackage{amssymb,amsmath}
\bibpunct{(}{)}{;}{a}{}{,}

\usepackage[pagebackref=true]{hyperref}
\hypersetup{colorlinks = true, linkcolor = green, anchorcolor = red, citecolor = blue, filecolor = red, urlcolor = red}

\begin{document}

   \title{Investigating the relationship between cosmic curvature and dark energy models with the latest supernova sample
}

   \volnopage{Vol.0 (20xx) No.0, 000--000}      
   \setcounter{page}{1}          

   \author{Chao Gao
      \inst{1}
   \and Yun Chen
      \inst{2}
   \and Jie Zheng
      \inst{1}
   }

   \institute{Department of Astronomy of Beijing Normal University,
             Beijing 100089, China; {\it gaoch@mail.bnu.edu.cn}\\
        \and
             Key Laboratory for Computational Astrophysics, National Astronomical Observatories, Chinese Academy of Sciences, Beijing, 100101, China;  {\it chenyun@nao.cas.cn}\\
\vs\no
   {\small Received~~20xx month day; accepted~~20xx~~month day}}

\abstract{We investigate the relationship between the cosmic curvature and the model of dark energy (hereafter DE) with the recent Type Ia supernovae (hereafter SNe Ia) data, i.e.,
the Pantheon sample including 1048 SNe Ia with $0.01 < z < 2.3$. We obtain the measurements of the dimensionless spatial curvature density today, i.e., $\Omega_{k0} = -0.062^{+0.189}_{-0.169}, -0.004^{+0.228}_{-0.134}, 0.127^{+0.280}_{-0.276}$ and $0.422^{+0.213}_{-0.338}$ at 68\% confidence level (CL), respectively, in the scenarios of $\Lambda$CDM, $\phi$CDM (i.e., scalar field dark energy), $\omega$CDM and $\omega_0\omega_a$CDM models. In the scenario of $\Lambda$CDM model, a closed universe is preferred by the Pantheon sample, which is consistent with that from the Planck CMB spectra. However, the uncertainty of $\Omega_{k0}$ from the Pantheon SNe sample is about 8 times larger than that from the Planck data, so the former one supports a closed universe at a much lower CL than that from the latter one.  An open unverse is supported by the Pantheon sample at $\sim$32\% and $\sim$78\% CLs, respectively, in the $\omega$CDM and $\omega_0\omega_a$CDM models.  Among these models, the $\phi$CDM model is the one which supports the flat universe most strongly. It shows that $\Omega_{k0}$ is significantly dependent on the adopted model of dark energy, and there is a negative correlation between $\Omega_{k0}$ and the equation of state of DE.
}

   \authorrunning{Chao Gao, Yun Chen \& Jie Zheng }            
   \titlerunning{cosmic curvature and dark energy models}  

   \maketitle

%
%
\section{Introduction}           
\label{sect:intro}

As a kind of ``standard candles'' in the universe, the SNe Ia supplied the first straightforward proof for the accelerating universe and for the existence of the unknown ``dark energy'' driving this acceleration in 1998. At that time, the sample size was not big, i.e., 50 SNe Ia from \citet{Riess+etal+1998}, and 42 ones from \citet{Perlmutter+etal+1999}. The population of SNe Ia discovered has been growing rapidly over the last two decades. The popular samples include the ``gold'' 2004 (157 data; \citealt{Riess+etal+2004}) and ``gold'' 2007 (182 data; \citealt{Riess+etal+2007}) samples,  the Supernova Legacy Survey (SNLS) 1-year (115 data; \citealt{Astier+etal+2006}) and 3-year (252 data; \citealt{Guy+etal+2010}) samples, the ESSENCE (``Equation of State: SupErNovae
trace Cosmic Expansion'' supernova survey) sample (60 data; \citealt{Miknaitis+etal+2007, Wood-Vasey+etal+2007,  Davis+etal+2007}), the Supernova Cosmology Project (SCP) Union (307 data; \citealt{Kowalski+etal+2008}), Union2 (557 data; \citealt{Amanullah+etal+2010}) and Union2.1 (580 data; \citealt{Suzuki+etal+2012}) compilations, the constitution set (397 data; \citealt{Hicken+etal+2009}), the ``joint light-curve analysis'' (JLA) compilation (740 data; Betoule et al. 2014), and the latest ``Pantheon'' sample (1048 data; \citealt{Scolnic+etal+2018}). Besides the dramatic increase of the population of SNe Ia, the techniques for measuring the light curve parameters are also continually being improved to reduce the systematic uncertainties (\citealt{Riess+etal+1996, Perlmutter+etal+1997,  Tonry+etal+2003, Wang+etal+2003, Wang+etal+2006, Guy+etal+2005, Guy+etal+2007, Conley+etal+2008}). At present, the most popular techniques mainly include the SALT/SALT2(\citealt{Guy+etal+2005,Guy+etal+2007}) and SiFTO (\citealt{Conley+etal+2008}) models which fit the light curves of supernovae by employing a spectral template.

The Cosmic Microwave Background (CMB) as one of the standard cosmological probes has revealed a strong evidence (i.e. at more than 99\%CL) for a closed universe in the non-flat $\Lambda$CDM model, by using the near-term Planck CMB spectra (\citealt{Aghanim+etal+2018, Aghanim+etal+2019, Di+Valentino+etal+2020}). The observational constraints on the cosmic curvature are widely studied with different probes (\citealt{Gong+etal+2008, Liao+etal+2017, Wang+etal+2017, Denissenya+etal+2018, Cao+etal+2019, Liao+2019, Qi+etal+2019, Wei+Melia+2020, Zhou+Li+2020}).  In this work, we intend to explore what type of cosmic curvature another standard cosmological probe, i.e., SNe Ia, may support. In our analysis, the SNe Ia dataset adopted is the Pantheon sample including 1048 data with $0.01 < z < 2.3$ (\citealt{Scolnic+etal+2018}).  We also focus on investigating the relationship between the cosmic curvature and the DE model. In practice, four cosmological models with different kinds of equation of state (EoS) of DE are taken into account. They are the $\Lambda$CDM model with the cosmological constant owning an EoS $\omega = -1$ (\citealt{Peebles+1984}), the $\phi$CDM model with the scalar field DE
owning a time-varying EoS $-1 < \omega < 0$ (\citealt{Peebles+Ratra+1988}), the $\omega$CDM model with the phenomenological DE owning an EoS $\omega = \textrm{Constant}$ (\citealt{Ratra+1991}), and the
$\omega_0\omega_a$CDM model with the dynamical DE owning a parameterized EoS $\omega(z) = \omega_0+\omega_a\frac{z}{1+z}$ proposed in \citet{Chevallier+Polarski+2001} and \citet{Linder+2003}.

The paper is organized as follows: in Section 2, we present the cosmological models under consideration, and demonstrate the methodology of using the SNe Ia data to put constraits on the  model parameters. In Section 3, we carry out observational constraints on the effective energy density of the cosmic curvature $\Omega_{k0}$, and other parameters in the considered cosmological models, and then mainly analyze the relationships between $\Omega_{k0}$ and the EoS of DE. The main conclusions and discussions are summarized in the last section .



\section{Methodology and dark energy Models}
\label{sect:Obs}

To put constraints on the cosmological parameters with the SNe Ia sample, one first needs to have the Friedmann equations for the cosmological models under consideration. According to the scope of this paper, the cosmic curvature, parameterized through the effective energy density parameter $\Omega_{k0}$ is taken to be a free parameter, rather than zero.

 Among the various types of cosmological models, the most economical one may be the $\Lambda$CDM model (\citealt{Peebles+1984}), in which the accelerating expansion of the universe is powered by the DE component modeled as the Einstein's cosmological constant, $\Lambda$, with an EoS parameter $\omega = p_\Lambda/\rho_\Lambda = -1$, where $p_\Lambda$ and $\rho_\Lambda$ are the fluid pressure and energy density. The Friedmann equation of the $\Lambda$CDM model is
 \begin{equation}
\label{eq:LCDMFriedmann} E^2(z;\textbf{p}) =
\Omega_{m0}(1+z)^3+\Omega_{\Lambda}+\Omega_{k0}(1+z)^2,
\end{equation}
where $E(z) = H(z)/H_0$ is the reduced
Hubble parameter defined with the Hubble parameter $H(z)$ and the Hubble constant $H_0 = H(z=0)$. The model parameters are $\textbf{p} = (\Omega_{m0},
\Omega_{k0})$,  where $\Omega_{m0}$ is the matter density parameter, $\Omega_{k0}$ is the effective energy density parameter of the curvature, and $\Omega_{\Lambda} = 1-\Omega_{m0}-\Omega_{k0}$ is the energy density parameter of $\Lambda$. In this paper,
we use the subscript $0$ to denote the present-day value of a quantity.

In the $\phi$CDM model, the DE is treated as the scalar field $\phi$ with a potential-energy density $V(\phi)$ decreasing gradually in $\phi$, in which the DE density decreases slowly in time. For the scalar field DE, several kinds of $V(\phi)$ can satisfy the requirement of the late-time accelerating expansion of the universe (\citealt{Samushia+2009}).  We consider the scalar field DE with a potential-energy density $V(\phi) = \frac{1}{2} \kappa m_p^2 \phi^{-\alpha}$, where $m_p = 1/\sqrt{G}$ is the Planck mass and $G$ is the Newtonian constant of gravitation, $\alpha$ and $\kappa$ are constants which should be greater than or equal to zero (\citealt{Ratra+Peebles+1988}). The $\phi$CDM model under consideration has been extensively studied (\citealt{Samushia+etal+2010, Chen+and+Ratra+2011, Chen+and+Ratra+2012, Mania+Ratra+2012, Chen+and+Xu+2016, Chen+etal+2015, Chen+etal+2016, Chen+etal+2017, Farooq+etal+2017, Ryan+etal+2019}). It can reduce to the $\Lambda$CDM model in the case of taking $\alpha = 0$. The Friedmann equation of this model is
\begin{equation}
\label{eq:phiCDMFriedmann} H^2(z;\textbf{p}) =
\frac{8\pi}{3m_p^2}(\rho_m + \rho_{\phi})-\frac{k}{a^2},
\end{equation}
where the Hubble parameter is defined as $H(z) = \dot{a}/a$, and $a(t)$ is the
cosmic scale factor and $\dot{a} = \textit{d}a/\textit{d}t$.
The DE energy density is
\begin{equation}
\label{eq:rhophi} \rho_{\phi} = \frac{m_p^2}{16\pi} (\frac{1}{2}\dot{\phi}^2 + V(\phi)).
\end{equation}
The EoS is
\begin{equation}
\label{eq:EoS_phi} \omega = \frac{\frac{1}{2}\dot{\phi}^2 - V(\phi)}{\frac{1}{2}\dot{\phi}^2 + V(\phi)}.
\end{equation}
One can figure out that this EoS satisfies $-1 < \omega < 1$.
The motion equation for $\phi$ can be expressed as
\begin{equation}
\label{eq:dotphi} \ddot{\phi} + 3 \frac{\dot{a}}{a}\dot{\phi} + \frac{\partial V(\phi)}{\partial \phi}= 0.
\end{equation}
The Hubble parameter $H(z)$ can be computed numerically with Eqs.\ (\ref{eq:phiCDMFriedmann}) and (\ref{eq:dotphi}),  as well as the initial conditions described in Peebles \& Ratra (1988). According to the usual convention, the effective energy density of the spatial curvature $k$ is defined as $\Omega_k(a) \equiv -k/(a^2H(z)^2)$, so its present-day value is $\Omega_{k0}=\Omega_k(z=0)=-k/(a_0^2H_0^2)$. In the $\phi$CDM model, the model parameters are $\textbf{p} = (\Omega_{m0}, \Omega_{k0}, \alpha)$.

In the $\omega$CDM model, the EoS of DE is taken as $\omega = \textrm{Constant}$. It reduces to the $\Lambda$CDM model in the case of taking $\omega = -1$. One can obtain the Friedmann equation
\begin{equation}
\label{eq:wCDMFriedmann} E^2(z;\textbf{p}) =
\Omega_{m0}(1+z)^3+(1-\Omega_{m0}-\Omega_{k0})(1+z)^{3(1+\omega)}+\Omega_{k0}(1+z)^2,
\end{equation}
where the model parameters are $\textbf{p} = (\Omega_{m0}, \Omega_{k0}, \omega)$.

The $\omega_0\omega_a$CDM model can be deemed as an extension of the $\Lambda$CDM  and $\omega$CDM models, in which the DE is modeled as a dynamical component with the EoS parameterized as $\omega = \omega_0 + \omega_a z/(1+z)$. It reduces to the $\Lambda$CDM model in the case of taking $\omega_0 = -1$ and $\omega_a = 0$, and to $\omega$CDM model in the case of taking $\omega_0 = \textrm{Constant}$ and $\omega_a = 0$.  Obviously, $\omega_a$ is a key parameter to denote the dynamic level of the DE. The Friedmann equation of the $\omega_0\omega_a$CDM model satisfies
\begin{equation}\label{eq:w0waCDMFriedmann}
E^2(z;\textbf{p})=\Omega_{m0}(1+z)^3+\Omega_{de0}(1+z)^{3(1+w_0+w_a)}e^{-\frac{3w_az}{1+z}}+\Omega_{k0}(1+z)^2,
\end{equation}
where the present-day value of the DE energy density is $\Omega_{de0} = 1-\Omega_{m0}-\Omega_{k0}$.

To constrain the cosmological parameters with the SNe Ia data, one should first figure out the corresponding observable and its theoretical (predicted) value.  The observable given in the ``Pantheon'' data set is the corrected magnitude $m_{cor}$ (see Table A17 of Scolnic et al. 2018), i.e.,
\begin{equation}\label{eq:corr_mag}
m_{cor} \equiv m_B+K = \mu+M,
\end{equation}
where $\mu$ is the distance modulus, $m_B$ is the apparent B-band magnitude, and $M$ is the absolute B-band magnitude of a fiducial SN Ia. According to the Eq.(3) of Scolnic et al. (2018), we can get the correction term $K = \alpha x_1-\beta c+\Delta_M+\Delta_B$ which includes the corrections related to four different sources (for more details, see Scolnic et al. 2018).
According the the definition of the distance modulus, one has
\begin{equation}\label{dist_mod}
\mu = 5\log(d_L)+25,
\end{equation}
where $d_L$ is the luminosity distance in Mpc.
The observable $Y^{obs} = \mu+M$ displayed in Eq.(\ref{eq:corr_mag}) should correspond to the theoretical (predicted) value
\begin{eqnarray}
Y^{th}&=& 5\log(d_L)+25 +M \nonumber\\
&=&5\log[(1+z)D(z)]+ Y_0,
\label{eq:Y_th}
\end{eqnarray}
where the constant term $Y_0$ is written as $Y_0 = M+5log(\frac{cH_0^{-1}}{Mpc})+25$, and the normalized comoving distance $D(z)$ is defined by,
\begin{equation}\label{eq:D_z}
d_L(z) = \frac{c(1 + z)}{H_0}D(z),
\end{equation}
where $c$ is the velocity of light. The normalized comoving distance $D(z)$ can be expressed as
\begin{equation}\label{eq:dl_z}
D(z)=\left\{
\begin{aligned}
&\frac{1}{\sqrt{-\Omega_{k0}}} \sin\left(\sqrt{-\Omega_{k0}}\int_0^z\frac{d\tilde{z}}{E(\tilde{z})}\right)\;\;\;\;\;\;\textrm{if}\;\;\Omega_{k0}<0,\\
&\int_0^z\frac{d\tilde{z}}{E(\tilde{z})}\ \ \ \ \ \ \ \ \ \ \ \ \ \ \ \ \ \ \ \ \ \ \ \ \ \ \;\;\;\;\;\;\;\;\;\;\;\;\;\;\;\;\;\;\textrm{if}\;\;\Omega_{k0}=0,\\
&\frac{1}{\sqrt{\Omega_{k0}}} \sinh\left(\sqrt{\Omega_{k0}}\int_0^z\frac{d\tilde{z}}{E(\tilde{z})}\right)\;\;\;\;\;\;\;\;\;\textrm{if}\;\;\Omega_{k0}>0,
\end{aligned} \right.
\end{equation}

The likelihood of Pantheon sample is given by
\begin{equation}
\label{eq:likelihood}
\mathcal{L}\propto e^{-\chi^2/2}.
\end{equation}
$\chi^2$ is constructed as
\begin{equation}
\label{eq:chi2}
\chi^2={\Delta \overrightarrow{Y}}^T\cdot\textbf{C}^{-1}\cdot{\Delta \overrightarrow{Y}},
\end{equation}
where the residual vector for the SNe Ia data in the Pantheon sample is $\Delta \overrightarrow{Y}_i = [Y^{obs}_i-Y^{th}(z_i; Y_0,\textbf{p})]$. The covariance matrix $\textbf{C}$ of the sample includes the contributions from both the statistical and systematic errors. The nuisance parameter, i.e., the constant term $Y_0$  is marginalized over with the analytical methodology presented in \citet{Giostri+etal+2012}. The posterior probability distributions of model parameters are obtained with an affine--invariant Markov chain Monte Carlo (MCMC) Ensemble sampler (emcee; \citealt{Foreman-Mackey+etal+2013}), where the likelihood can be worked out with Eqs. (\ref{eq:likelihood}) and (\ref{eq:chi2}). We assume a flat prior for each parameter over a range of interest. In the framework of each cosmological model, the number of walkers is set as the number of the model parameters times 40, and the number of steps is 3000.


\section{Analysis and results}
\label{sect:Ana}

In the frameworks of the cosmological models under consideration, the observational constraints from the Pantheon sample are presented in Table \ref{tab:results}, including the mean values and 68\% confidence limits of the parameters.
In the $\Lambda$CDM model, a closed universe is preferred with a mean value $\Omega_{k0} = -0.062$, but at a non-high CL ($\sim$ 25\% CL) because of a high uncertainty. The result is consistent with that from \citet{Wang+2018}, in which the non-flat $\Lambda$CDM model is constrained with the Pantheon sample via the MCMC code CosmoMC (\citealt{Lewis+2013}). In the $\phi$CDM model, it prefers a flat universe with $\Omega_{k0} = -0.004^{+0.228}_{-0.134}$ at 68\% CL. An open universe is preferred in both the $\omega$CDM and $\omega_0\omega_a$CDM models, according to $\Omega_{k0} = 0.127^{+0.280}_{-0.276}$ and $0.422^{+0.213}_{-0.338}$ at 68\% CL, respectively. It turns out that the bound on $\Omega_{k0}$ is significantly dependent on the adopted dark energy model.  Further, we use the Bayesian information criterion (BIC) to do the model comparison. The BIC (\citealt{Schwarz+1978}) is defined as
\begin{equation}
\label{eq:BIC}
\textrm{BIC}=-2\ln \mathcal{L}_{max}+k\ln N,
\end{equation}
where $\mathcal{L}_{max}$ is the maximum likelihood (i.e., $-2\ln \mathcal{L}_{max}=\chi^2_{min}$ under the Gaussian assumption), $k$ is the number of the model parameters, and $N$ is the size of the sample used in the analysis.  The BIC is widely used in the cosmological context (see e.g. \citealt{Liddle+2004, Biesiada+2007, Li+etal+2013, Birrer+2019, Chen+etal+2019}). The favorite model should be the one with the minimum BIC value.
The BIC values for the $\Lambda$CDM, $\phi$CDM, $\omega$CDM and $\omega_0\omega_a$CDM models are 1040.6, 1047.4, 1047.3 and 1054.4, respectively. So, the $\Lambda$CDM model is the one which fits the Pantheon SNe sample best.
\begin{table*}
\begin{center}
\caption{\label{tab:results}  Observational constraints on the parameters of interest from the Pantheon SNe sample.
The mean values with 68\% confidence limits are displayed.}
\begin{tabular}{lllllll}\hline\hline
Models & Parameters &  &  & &$\chi^2_{\textrm{min}}/\textrm{d.o.f}$ & BIC   \\
\hline
$\Lambda$CDM & $\Omega_{k0} = -0.062^{+0.189}_{-0.169}$ & $\Omega_{m0} = 0.323^{+0.069}_{-0.075}$ & ... & ...& $1026.7/1048$ & 1040.6 \\
$\phi$CDM &$\Omega_{k0} = -0.004^{+0.228}_{-0.134}$ & $\Omega_{m0} = 0.215^{+0.060}_{-0.082}$ & $\alpha=0.679^{+1.168}_{-0.505}$ & ...& $1026.5/1048$ & 1047.4\\
$\omega$CDM & $\Omega_{k0}=0.127^{+0.280}_{-0.276}$ & $\Omega_{m0} = 0.288^{+0.076}_{-0.078}$ & $\omega=-1.236^{+0.346}_{-0.722}$ & ...& $1026.4/1048$ & 1047.3\\
$\omega_0\omega_a$CDM& $\Omega_{k0}=0.422^{+0.213}_{-0.338}$ & $\Omega_{m0} = 0.219^{+0.134}_{-0.129}$ & $\omega_0=-1.784^{+0.770}_{-1.291}$ & $\omega_a=-0.110^{+6.116}_{-5.289}$ & $1025.6/1048$ & 1054.4\\
\hline
\end{tabular}
\end{center}
\end{table*}

\begin{figure}
\begin{center}
\includegraphics[angle=0, width=0.9\hsize]{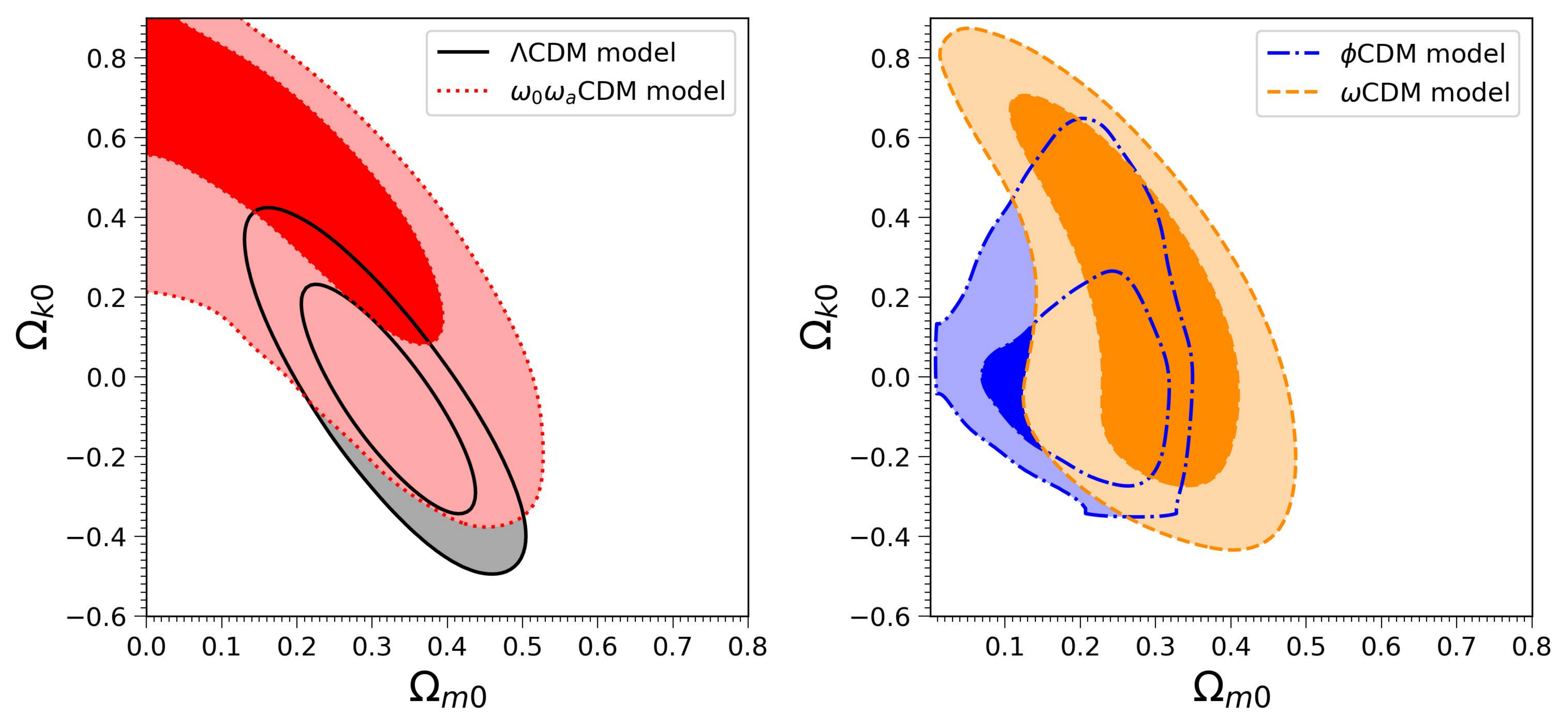}
\end{center}
\caption{Contours in the $(\Omega_{m0},\Omega_{k0})$ plane refer to the two-dimensional (2D) marginalized distributions at 68\% and 95\% CLs, constrained with the Pantheon sample in the scenarios of $\Lambda$CDM, $\phi$CDM, $\omega$CDM and $\omega_0\omega_a$CDM models, respectively.}
\label{fig:ok_om}
\end{figure}

\begin{figure}
\begin{center}
\includegraphics[angle=0, width=0.9\hsize]{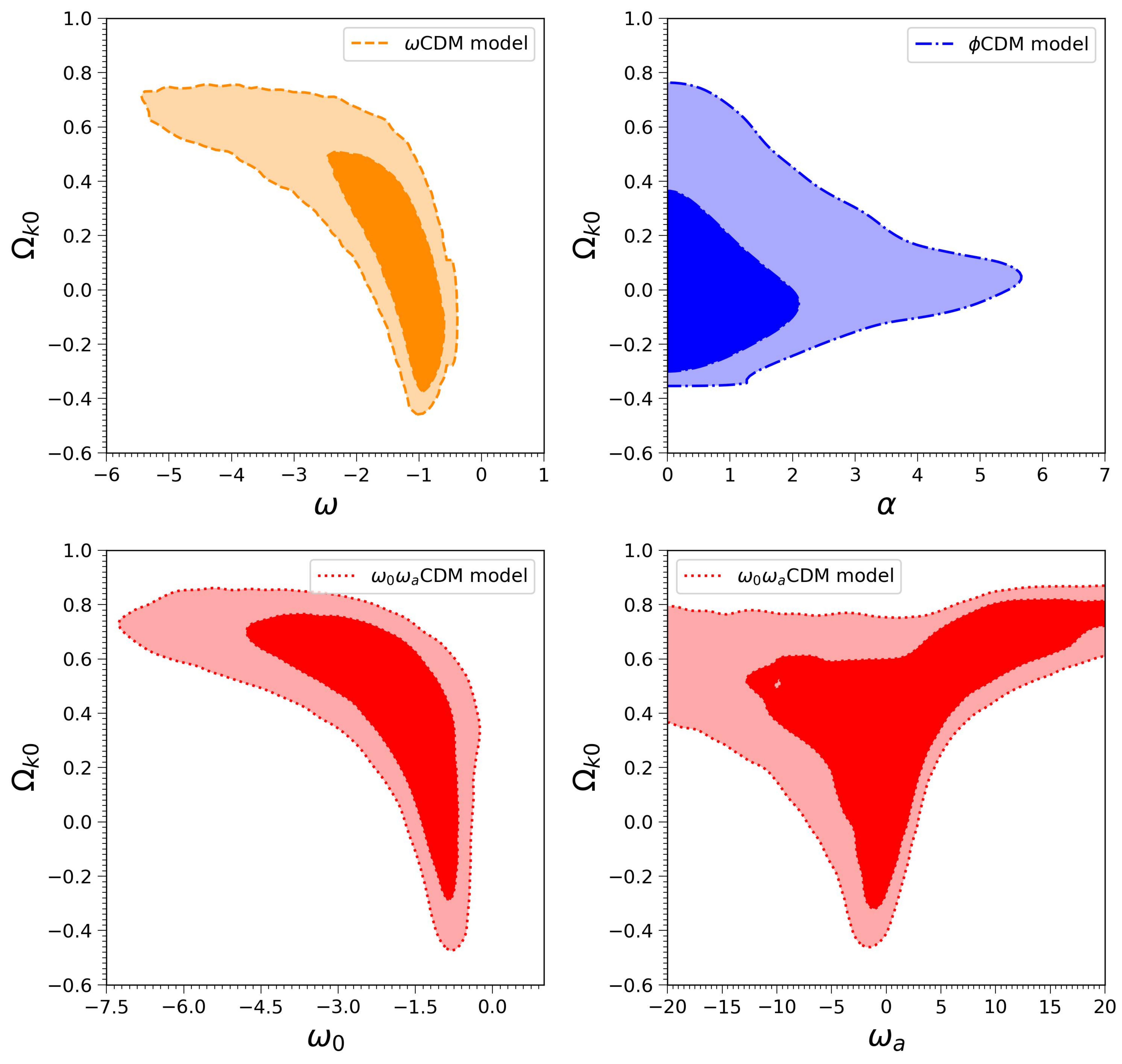}
\end{center}
\caption{The contours correspond to the 2D probability distributions at 68\% and 95\% confidence levels for parameters of interest.}
\label{fig:ok_others}
\end{figure}

 To study the correlation between $\Omega_{k0}$ and $\Omega_{m0}$, we display the two-dimensional (2D) probability distributions in the $(\Omega_{m0},\Omega_{k0})$ plane for all the cosmological models under consideration in Fig.\ref{fig:ok_om}. One can find the negative correlation between $\Omega_{k0}$ and $\Omega_{m0}$ in $\Lambda$CDM, $\omega$CDM and $\omega_0\omega_a$CDM scenarios. Besides, there is not an apparent correlation between them in the $\phi$CDM scenario. Then, we turn to study the relations between $\Omega_{k0}$ and other parameters besides $\Omega_{m0}$ in Fig.\ref{fig:ok_others}. We find a negative correlation between $\Omega_{k0}$ and the DE EoS in the $\omega$CDM model from the upper-left panel of Fig.\ref{fig:ok_others}. The upper-right panel of Fig.\ref{fig:ok_others} displays that there is not an obvious correlation between $\Omega_{k0}$ and $\alpha$ in the $\phi$CDM scenario. From the lower panels of Fig.\ref{fig:ok_others}, we find a negative correlation between $\Omega_{k0}$ and $\omega_0$, but no obvious correlation between $\Omega_{k0}$ and $\omega_a$ is discovered in the $\omega_0\omega_a$CDM model.

In the $\Lambda$CDM scenario, the mean value $\Omega_{k0} = -0.062$ constrained from the Pantheon SNe sample is close to but a bit smaller than the one $\Omega_{k0} = -0.044$ from the Planck CMB spectra (\citealt{Aghanim+etal+2018}). Nevertheless, the uncertainty of $\Omega_{k0}$ from the Pantheon sample is about 8 times larger than that from the Planck data, hence the former supports a closed universe at a much lower CL (at $\sim$ 25\% CL) than that from the latter (at $\sim$ 99\% CL). Moreover, as discussed in Di Valentino (2020), when jointing the Planck CMB along with the BAO data, $\Omega_{k0}$ increases to $\Omega_{k0} = 0.0008^{+0.0038}_{-0.0037}$ at 95\% CL(\citealt{Aghanim+etal+2018, Di+Valentino+etal+2020}). It turns out that the limit on $\Omega_{k0}$ changes significantly with the data sets adopted.


\section{Conclusions}
\label{sect:conclusion}

By considering four different kinds of DE models, we have studied the relation between the energy density of spatial curvature $\Omega_{k0}$ and the DE model with the recent SNe Ia data, i.e., the Pantheon sample.  It turns out that the bound on $\Omega_{k0}$ is dependent on the adopted DE model notably, and a negative correlation exists between $\Omega_{k0}$ and the DE equation of state. Briefly speaking, a closed universe is preferred in the $\Lambda$CDM model;  a flat universe is heavily supported in the $\phi$CDM model; a open universe is favored in the $\omega$CDM and $\omega_0\omega_a$CDM models.

In the scenario of $\Lambda$CDM model, the limits on $\Omega_{k0}$ at 68\% are $\Omega_{k0} = -0.062^{+0.189}_{-0.169}$ from the Pantheon sample, and $\Omega_{k0} = -0.044^{+0.018}_{-0.015}$ from the Planck CMB spectra (\citealt{Aghanim+etal+2019}). Both the Pantheon SNe sample and the Planck CMB spectra data  support a closed universe. Nevertheless, the uncertainty of $\Omega_{k0}$ from the former one is much larger than that from the latter one, thus the former one supports a closed universe at a much lower CL (at $\sim$ 25\% CL) than that from the latter one (at $\sim$ 99\% CL). In addition, when combining the Planck CMB with the BAO data, the value of $\Omega_{k0}$ changes to $\Omega_{k0} = 0.0008^{+0.0038}_{-0.0037}$ at 95\% CL, that is in good agreement with a flat universe. It reflects the sample dependence of the limit on $\Omega_{k0}$. Consequently, in view of the noticeable model-dependence and sample-dependence of the limit on $\Omega_{k0}$, one should modestly use the assumption of a flat universe.

\begin{acknowledgements}
This work has been supported by the National Natural Science Foundation of China (Nos. 11633001, 11920101003, 11703034, 11773032 and 11573031), the Strategic Priority Research Program of the Chinese Academy of Sciences (No. XDB23000000), the Interdiscipline Research Funds of Beijing Normal University, and the NAOC Nebula Talents Program.
\end{acknowledgements}

\label{lastpage}


\begin{thebibliography}{}

\bibitem[Aghanim et al. (2018)]{Aghanim+etal+2018}Aghanim, N., Akrami, Y., Ashdown, M., et al. [Planck Collaboration], 2018, arxiv e-prints, arXiv:1807.06209
\bibitem[Aghanim et al. (2019)]{Aghanim+etal+2019}Aghanim, N., Akrami, Y., Ashdown, M., et al. [Planck Collaboration], 2019, arxiv e-prints, arXiv:1907.12875

\bibitem[Amanullah et al. (2010)]{Amanullah+etal+2010}Amanullah, R., Lidman, C., Rubin, D., et al. 2010, ApJ, 716, 712

\bibitem[Astier et al. (2006)]{Astier+etal+2006}Astier, P., Guy, J., Regnault, N., et al., 2006, A\&A, 447, 31

\bibitem[Betoule et al. (2014)]{Betoule+etal+2014}Betoule, M., Kessler, R., Guy, J., et al. 2014, A\&A, 568, A22

\bibitem[Biesiada (2007)]{Biesiada+2007}Biesiada, M. 2007, J. Cosmol. Astropart. Phys., 2, 3

\bibitem[Birrer (2019)]{Birrer+2019}Birrer, S., Treu, T., Rusu, C. E., et al., 2019, MNRAS, 484, 4726

\bibitem[Cao et al. (2019)]{Cao+etal+2019} Cao, S., Qi, J., Biesiada, M., et al., 2019, Physics of the Dark Universe, 24, 100274

\bibitem[Chen \& Ratra (2011)]{Chen+and+Ratra+2011}Chen, Y., \& Ratra, B.  2011, PhLB, 703, 406

\bibitem[Chen \& Ratra (2012)]{Chen+and+Ratra+2012}Chen, Y., \& Ratra, B.  2012, A\&A 543, A104

\bibitem[Chen et al. (2015)]{Chen+etal+2015}Chen, Y., Geng, C.-Q., Cao, S., et al. 2015, JCAP, 02, 010

\bibitem[Chen \& Xu (2016)]{Chen+and+Xu+2016}Chen, Y., \& Xu, L. 2016, PhLB, 752, 66

\bibitem[Chen et al. (2016)]{Chen+etal+2016}Chen, Y., Ratra, B., Biesiada, M., et al. 2016, ApJ, 829, 61

\bibitem[Chen et al. (2017)]{Chen+etal+2017}Chen, Y., Kumar, S.,  \& Ratra, B. 2017, ApJ, 835, 86

\bibitem[Chen et al. (2019)]{Chen+etal+2019}Chen, Y., Li, R., Shu, Y., \& Cao, X. 2019, MNRAS, 488, 3745

\bibitem[Chevallier \& Polarski (2001)]{Chevallier+Polarski+2001}Chevallier, M. \& Polarski, D., 2001, Int. J. Mod. Phys. D, 10, 213

\bibitem[Conley et al. (2008)]{Conley+etal+2008}Conley, A., Sullivan, M., Hsiao, E. Y., et al., 2008, ApJ, 681, 482

\bibitem[Davis et al. (2007)]{Davis+etal+2007}Davis, T. M., M\''{o}rtsell, E., Sollerman, J., et al., 2007, ApJ, 666, 716

\bibitem[Denissenya et al. (2018)]{Denissenya+etal+2018}Denissenya, M., Linder, E. V., \& Shafieloo, A., 2018, JCAP, 03, 041

\bibitem[Di Valentino et al. (2020)]{Di+Valentino+etal+2020}Di Valentino, E., Melchiorri, A., Silk, J., 2020, Nature Astronomy, 4, 196

\bibitem[Foreman-Mackey et al. (2013)]{Foreman-Mackey+etal+2013}Foreman-Mackey, D., Hogg, D.W., Lang, D., \& Goodman, J., 2013, PASP, 125, 306

\bibitem[Giostri et al. (2012)]{Giostri+etal+2012}Giostri, R., Vargas dos Santos, M., Waga, I., et al., 2012, JCAP, 03, 027

\bibitem[Gong et al. (2008)]{Gong+etal+2008}Gong, Y., Wu, Q., \& Wang, A., 2008, ApJ, 681, 27

\bibitem[Guy et al. (2005)]{Guy+etal+2005}Guy, J., Astier, P., Nobili, S., Regnault, N., \& Pain, R. 2005, A\&A, 443, 781

\bibitem[Guy et al. (2007)]{Guy+etal+2007}Guy, J., Astier, P., Baumont, S., et al. 2007, A\&A, 466, 11

\bibitem[Guy et al. (2010)]{Guy+etal+2010}Guy, J., Sullivan, M., Conley, A., et al., 2010, A\&A, 523, A7

\bibitem[Hicken et al. (2009)]{Hicken+etal+2009} Hicken, M., Wood-Vasey, W. M., Blondin, S., et al., 2009, \apj, 700, 1097

\bibitem[Kowalski et al. (2008)]{Kowalski+etal+2008}Kowalski, M., Rubin, D., Aldering, G., et al., 2008, ApJ, 686, 749

\bibitem[Lewis (2013)]{Lewis+2013}Lewis, A., 2013, Phys. Rev. D, 87, 103529

\bibitem[Li et al. (2013)]{Li+etal+2013}Li, M., Li, X.-D., Wang, S., \& Wang, Y., 2013, Front. Phys., 8, 828

\bibitem[Liao (2019)]{Liao+2019}Liao, K., 2019, Phys Rev D, 99, 083514

\bibitem[Liao et al. (2017)]{Liao+etal+2017}Liao, K., Li, Z., Wang, G., \& Fan, X.,  2017, ApJ, 839, 70

\bibitem[Liddle (2004)]{Liddle+2004}Liddle, A. R., 2004, MNRAS, 351, L49

\bibitem[Linder (2003)]{Linder+2003}Linder, E. V., 2003, Phys. Rev. Lett., 90, 091301

\bibitem[Mania \& Ratra (2012)]{Mania+Ratra+2012}Mania, D., \& Ratra, B.,  2012, PhLB, 715,9

\bibitem[Miknaitis et al. (2007)]{Miknaitis+etal+2007} Miknaitis, G., Pignata, G., Rest, A., et al., 2007, ApJ, 666, 674

\bibitem[Farooq et al. (2017)]{Farooq+etal+2017}Farooq, O., Ranjeet Madiyar, F., Crandall, S., \& Ratra, B. 2017, ApJ, 835, 26

\bibitem[Peebles (1984)]{Peebles+1984}Peebles, P. J. E., 1984, ApJ, 284, 439

\bibitem[Peebles \& Ratra (1988)]{Peebles+Ratra+1988}Peebles, P. J. E., \& Ratra, B., 1988, ApJ, 325, L17

\bibitem[Perlmutter et al. (1997)]{Perlmutter+etal+1997}Perlmutter, S., Gabi, S., Goldhaber, G., et al. 1997, ApJ, 483, 565

\bibitem[Perlmutter et al. (1999)]{Perlmutter+etal+1999}Perlmutter, S., Aldering, G., Goldhaber, G., et al. 1999, ApJ, 517, 565

\bibitem[Qi et al. (2019)]{Qi+etal+2019}Qi, J., Cao, S., Zhang, S., et al., 2019, MNRAS, 483, 1104

\bibitem[Ratra \& Peebles (1988)]{Ratra+Peebles+1988}Ratra, B., \& Peebles, P. J. E. 1988, Phys. Rev. D, 37, 3406

\bibitem[Ratra (1991)]{Ratra+1991}Ratra, B. 1991, Phys. Rev. D, 43, 3802

\bibitem[Riess et al. (1996)]{Riess+etal+1996}Riess, A. G., Press, W. H., \& Kirshner, R. P. 1996, ApJ, 473, 88

\bibitem[Riess et al. (1998)]{Riess+etal+1998} Riess, A. G., Filippenko, A. V., Challis, P., et al., 1998, AJ, 116, 1009

\bibitem[Riess et al. (2004)]{Riess+etal+2004}Riess, A. G., Strolger, L.-G., Tonry, J., et al., 2004, ApJ, 607, 665

\bibitem[Riess et al. (2007)]{Riess+etal+2007}Riess, A. G., Strolger, L.-G., Casertano, S., et al., 2007, ApJ, 659, 98

\bibitem[Ryan et al. (2019)]{Ryan+etal+2019}Ryan, J., Chen, Y, \& Ratra, B. 2019, MNRAS, 488, 3844

\bibitem[Samushia (2009)]{Samushia+2009}Samushia, L., 2009, arxiv e-prints, arXiv:0908.4597

\bibitem[Samushia et al. (2010)]{Samushia+etal+2010}Samushia, L., Dev, A., Jain, D.,  \& Ratra, B., 2010, PhLB, 693 509

\bibitem[Schwarz et al. (1978)]{Schwarz+1978}Schwarz, G., 1978, Ann. Stat., 6, 461

\bibitem[Scolnic et al. (2018)]{Scolnic+etal+2018}Scolnic, D. M., Jones, D. O., Rest, A., et al., 2018, ApJ, 859, 101

\bibitem[Suzuki et al. (2012)]{Suzuki+etal+2012}Suzuki, N., Rubin, D., Lidman, C., et al., 2012, ApJ, 746, 85

\bibitem[Tonry et al. (2003)]{Tonry+etal+2003}Tonry, J. L., Schmidt, B. P., Barris, B., et al. 2003, ApJ, 594, 1

\bibitem[Wang (2018)]{Wang+2018} Wang, D., 2018, Phys. Rev. D, 97, 123507

\bibitem[Wang et al. (2017)]{Wang+etal+2017}Wang, G., Wei, J., Li, Z., et al., 2017, ApJ, 847, 45

\bibitem[Wang et al. (2003)]{Wang+etal+2003}Wang, L., Goldhaber, G., Aldering, G., \& Perlmutter, S. 2003, ApJ, 590, 944

\bibitem[Wang et al. (2006)]{Wang+etal+2006}Wang, L., Strovink, M., Conley, A., et al. 2006, ApJ, 641, 50

\bibitem[Wei \& Melia (2020)]{Wei+Melia+2020}Wei, J.,\& Melia, F., 2020, ApJ, 888, 99

\bibitem[Wood-Vasey et al. (2007)]{Wood-Vasey+etal+2007}Wood-Vasey, W. M., Miknaitis, G., Stubbs, C. W., et al., 2007, ApJ, 666, 694

\bibitem[Zhou \& Li (2020)]{Zhou+Li+2020}Zhou, H. \& Li, Z., 2020, ApJ, 889, 186



\end{thebibliography}
\end{document}